\documentclass[letterpaper, 10 pt, conference]{ieeeconf}  

\IEEEoverridecommandlockouts                              

\usepackage[latin1]{inputenc}
\usepackage{amsmath}
\usepackage{amsfonts}
\usepackage{amssymb}
\usepackage{color}

\usepackage{cite}

\usepackage{graphicx}
\graphicspath{{Fig/}{../Fig/}}

\usepackage{textcomp}
\usepackage{amsmath, amssymb}
\usepackage{indentfirst}
\usepackage{color}
\definecolor{red}{rgb}{1,0,0}

\definecolor{blu}{rgb}{0,0,1}

\definecolor{magenta}{cmyk}{0.1, 1, 0, 0}
\definecolor{greenedu}{cmyk}{1, 0, 1, 0.1}
\definecolor{cyanedu}{cmyk}{1, 0, 0, 0.1}

\newcommand{\vecx}{\mathbf{x}}
\newcommand{\bydef}{\stackrel{\Delta}{=}}
\newcommand{\almosteq}{\stackrel{\sim}{=}}
\newcommand{\cqfd}{\hfill \rule{2mm}{2mm}\medbreak\indent}

\newtheorem{definition}{\bf Definition}

\begin{document}

\title{Learning about passivity from data}

\author{Alexandre Sanfelici Bazanella\thanks{A. S. Bazanella is with the Department of Automation and Energy, Universidade Federal do Rio Grande do Sul (DELAE/UFRGS), Porto Alegre-RS, Brazil. Email: bazanella@ufrgs.br. This study was financed by Conselho Nacional de Desenvolvimento Cient\'{\i}fico e Tecnol\'ogico (CNPq)}}

\maketitle
\thispagestyle{empty}
\pagestyle{empty}

\begin{abstract}
This paper presents a data-driven methodology to estimate the storage function of a passive system.
The methodology consists in parametrizing 
the storage function with a dictionary then running a linear program. Results on a benchmark are presented to
illustrate its properties, including its robustness to noise. Various uses of the storage function
that do not require knowledge of a model are also discussed.
\end{abstract}

\section{Introduction}

Passivity is a fundamental property of dynamical systems that plays a major role in systems and control theory at least since the 1960's. 
Determining whether a system is passive or not, and to which degree, revolves arounds finding an appropriate function describing its
energy dissipation properties - the storage function. Passivity concerns the input-output relation of the system, and yet
a storage function can also be used as a Lyapunov function, thus allowing to study the stability of the unforced behavior of the system
as well.  Once a storage function is known, various tasks can be performed using its  knowledge: estimate the domain of attraction, 
estimate the convergence rate of autonomous trajectories, design a controller to improve this convergence rate,
certify a previously designed controller, etc; this is all standard theory \cite{Khalil, Rodolphe}. 
What is much less recognized in the model-based tradition of systems and control theory is that these tasks
require knowledge only of the storage function and not of the model.

A storage function for a given system is usually obtained analytically from its model,
but  knowledge of a model is by no means a guarantee that a good storage function can be found.
There are no general fail-proof methods for that, except for linear systems.
Known methods are either for restricted classes of systems \cite{Khalil}, rely on nonrobust numerical procedures
\cite{Zubov, maximal_Lyapunov}, and/or are constructive design methods that actually passivate a given system through control, 
instead of verifying the passivity of the uncontrolled system \cite{Rodolphe,MiroslavPK}.
So, from an applications perspective the relationship between models and storage functions is somewhat problematic:
on one hand, in a variety of applications the model serves only as a means to find a storage function, becoming 
superfluous once it is found; but on the other hand the knowledge of a model is often not enough to find a useful storage function.
Therefore, alternative ways of finding a storage function that do not rely on an analytical model are to be welcomed, even when such a
model is available. 

This is what this paper is about: in it, a purely data-driven methodology to obtain a storage function is presented. 
The methodology consists in solving a linear optimization program (LP) formed with input/output/state data collected
from the system, without the need for any priors. The LP yields the optimal parameter values of a dictionary 
parametrization of the storage function. A hyperparameter of the LP allows to make the solution robust to noise and to numerical errors. 

Recently, research on data-driven control has experimented a boost \cite{meu_editorial, editorial_Florian}
and a considerable amount of this work involves Lyapunov analysis. 
Among these works, the ones  that deal with nonlinear systems usually consist of control synthesis
\cite{Italian_Job} (as opposed to analysis), and rarely deal with passivity \cite{Frank_ARC, Frank_Passivity}.
Also, the vast majority of this work treats discrete-time systems,
whereas this paper is devoted to continuous time systems.  

I start by reviewing in Section \ref{sec:passivity} the basic definitions of passivity. 
Then the proposed data-driven method for determination of passivity and identification of the storage function
will be presented in Section \ref{sec:formulation}. 
Results for a classical benchmark - the pendulum - are presented in Section \ref{sec:case}. 
Various uses of the storage function are discussed in Section \ref{sec:use}. 
Finally, lines of future and present work are outlined in Section \ref{sec:conclusion}.

\section{Passivity}\label{sec:passivity}

An overview of the theory on passivity of continuous time dynamical systems is given in this Section.
The presentation follows the fundamental reference \cite{Rodolphe}.
The class of systems considered are SISO (single-input-single-output) systems in the standard input-affine form:
\begin{eqnarray}
\dot{\mathbf{x}} & = &  f(\mathbf{x}) + g(\mathbf{x}) .u \label{state} \\
y & = & h(\mathbf{x}) \label{output}
\end{eqnarray}
where $\mathbf{x}\in\mathbb{R}^n$ is the state, $u\in\mathbb{R}$ and $y\in\mathbb{R}$ are the input and output respectively,
$f(\cdot ) : \mathbb{R}^n\rightarrow \mathbb{R}^n$ and $g(\cdot ) : \mathbb{R}^n\rightarrow \mathbb{R}^n$ are vector fields,
$f(\mathbf{0}) = \mathbf{0}$ - the origin is an equilibrium of the autonomous system - and
$h(\cdot ) : \mathbb{R}^n\rightarrow \mathbb{R}$ is the output function.  


\begin{definition} {\bf Dissipativity and Passivity. } 
Let $\omega(\cdot, \cdot): \mathbb{R}\times\mathbb{R}\rightarrow \mathbb{R}$ be an integrable function, 
called the {\em supply rate}.  A system in the form
\eqref{state}-\eqref{output} is said to be \underline{dissipative} with supply rate $\omega(u(t),y(t))$ if there
exists a function $S(\cdot) : \mathbb{R}^n\rightarrow \mathbb{R}$ satisfying,  in some domain 
of the state space containing the origin, the following properties:
\begin{eqnarray}
 S(\mathbf{0}) & = &  0 \label{zero} \\
 S(\mathbf{x}) &  \geq &  0  \label{V} \\
 S(\mathbf{x}(T)) - S(\mathbf{x}(0))   & \leq & \int_0^T \omega(u(t),y(t)) dt  \nonumber \\
&&  \forall u(t), \forall T>0 . \label{Vdot} 
 \end{eqnarray}
This function $S(\cdot)$ is called the {\em storage function}.

A system is said to be passive if it is dissipative  with supply rate $\omega(u(t),y(t))= u(t).y(t) $.  

If a system is dissipative  with supply rate $\omega(u(t),y(t)) = u(t).y(t) - \rho y^2(t)$ for some
$\rho \in \mathbb{R}$ then it is said to be output feedback passive and the notation $OFP(\rho)$ is used.

If a system is dissipative  with supply rate $\omega(u(t),y(t)) = u(t).y(t) - \nu u^2(t)$ for some
$\nu \in \mathbb{R}$ then it is said to be input feedback passive and the notation  $IFP(\nu)$ is used. 
\cqfd
\end{definition}

Positive values of $\rho$ and $\nu$ mean that the system has an excess of passivity; accordingly, negative
values mean that the system presents a shortage of passivity. 

Passivity being defined for all inputs $u(t)$, the definition is valid also  for $u(t)\equiv 0$. In this case \eqref{Vdot}
simplifies to $S(\mathbf{x}(T)) - S(\mathbf{x}(0))  \leq 0$, which implies $L_fS(\mathbf{x}) \leq 0$ - where
$L_fS(\mathbf{x}) \bydef \frac{\partial S(\vecx )}{\partial \vecx} f(\vecx )$ is the Lie derivative of $S(\vecx )$.  
Hence, a storage function is also a Lyapunov function if it is strictly positive\footnote{Note that passivity only requires semi-definiteness of $S(\cdot )$}
and in this case  the origin of a passive system is guaranteed to be asymptotically stable. 

The negative feedback connection of two passive systems results in a stable system. In particular,
the feedback $u=-ky$ results in a stable closed-loop system for any passive system and any $k>0$.
These observations also lead to a particular approach to the state feedback control design problem, 
which consists in looking for an output such that the system is passive. Given a Lyapunov function $S(\mathbf{x})$ for the system, 
if the output is picked as $h(\mathbf{x}) = L_g S(\mathbf{x})$ then the resulting system is passive and the feedback 
$u= - k.y = - k . L_g S(\mathbf{x})$ results in a stable closed-loop with improved damping with respect to the open-loop 
system for all $k>0$. Moreover, this control law is also optimal with respect to an LQ performance criterion. 
This design approach,  called $LgV$ control or damping control \cite{meu_LgV, Marta}, is a classical subject
in nonlinear system theory and has also inspired the development of  more ambitious design approaches - see \cite{energy_back_in_control},
for example. 

Storage/Lyapunov functions are not all the same. A good function is one that yields good results: a tight estimate
of the excess/shortage of passivity, an effective damping control law, a tight estimate of the domain of attraction of 
the equilibrium, etc. 
Finding good storage functions is an open problem in general, for which there are no fail-proof methods. 
This paper presents a solution to obtain just that - a storage function for an unknown system based on
experimental input-output-state data. This is accomplished by means of a linear optimization program,
as shown in the next Section.

\section{LP Formulation}\label{sec:formulation}

If the system under study is passive, then there exists a storage function $S(\cdot ):\mathbb{R}^n\rightarrow\mathbb{R}$. 
Let us assume that this function is sufficiently smooth, so that it can be described by means of a parametrization:
\begin{equation}\label{dictionary}
S(\mathbf{x}) = \theta^T \phi(\mathbf{x})
\end{equation}
where $\phi(\cdot): \mathbb{R}^n\rightarrow \mathbb{R}^d$ is a vector field, called the {\em dictionary}, and
$\theta\in\mathbb{R}^d$ is a parameter vector. To satisfy the constraint \eqref{zero}, it is convenient (though not necessary) to chose the
dictionary such that $\phi(\mathbf{0}) = \mathbf{0}$. Once a dictionary is chosen, determination of the storage function
boils down to determining an appropriate value of the parameter $\theta$. 

In order to obtain such an estimate of the parameter $\theta$, and thus of the storage function, one collects a set of data at $N$ time instants: 
$$
u(t) , \mathbf{x}(t), y(t) \quad t= t_1, t_2, \ldots t_N .
$$
If the system is passive, then the storage function must satisfy \eqref{V} and \eqref{Vdot}  at all data points, that is
\begin{equation}\label{qwer}
S(\mathbf{x}(t_i)) \geq 0 
\end{equation}
for all $t_i$ and 
\begin{equation}\label{asdf}
S(\mathbf{x}(t_j)) - S(\mathbf{x}(t_k))  \leq \int_{tk}^{tj} w(u(t), y(t)) ~dt
\end{equation}
for all pairs $t_j$, $t_k$ such that $t_j>t_k$.
The search for a storage function will consist of an optimization problem 
with the constraints \eqref{qwer} and \eqref{asdf}. 
For a set of $N$ data points, \eqref{qwer} results in $N$ constraints.

As for \eqref{asdf}, considering all  pairs
of time instants $t_k$ and $t_j$, there are $\sum_{i=1}^N (N-i) = O(N^2)$ such constraints.
However, it is unnecessary to write al of them as they will present high redundancy. 
On the other hand, noisy measurements may cause these constraints 
to be violated even with a correct storage function. So it is actually undesirable to include constraints with
$t_k$ and $t_j$ very close to each other, for which the effect of noise is most pronounced.
Accordingly, let us make use only of the pairs $t_j=t_{k+T}$ for some fixed
window size $T\in\mathbb{N}^+$, and all time instants for which this pair can be formed, that is,
for all $k=1, 2, \ldots , N-T$. In so doing the number of constraints is reduced to $N-T$ and the deleterious effect
of noise is mitigated for sufficiently large $T$.

It is also of interest to  estimate  the excess of passivity at the output or at the input. To do that, one can pick
either $\omega(u(t), y(t)) = u(t)y(t)-\rho y^2(t)$  or $\omega(u(t), y(t)) = u(t)y(t)-\nu u^2(t)$ as supply rate. 
In each case - IFP or OFP, that is - the equations will be slightly different, but the differences 
are rather obvious, so let me present only the case in which excess of passivity at the output is estimated. 
For this case, define an augmented parameter vector as
$$
\eta = \left[\begin{array}{c}
\theta \\ \rho 
\end{array}\right] ,
$$
and an augmented regressor as
$$
\varphi_a (t_k) = \left[\begin{array}{c}
\varphi( \mathbf{x}(t_k)) \\ \int_{tk}^{tj} y^2(t)dt 
\end{array}\right] .
$$
The identification of the storage function then  consists in the solution of the following optimization problem:
\begin{eqnarray}
&& \max_{\eta} \rho \label{cost} \\
&& s.t. \nonumber \\
&& \theta^T \varphi(\mathbf{x}(t_i)) > 0 ~~i=1, 2, \ldots , N\label{positive_inequality} \\
&& \eta^T ( \varphi_a (t_{j}) - \varphi_a (t_k))\leq \int_{tk}^{tj} u(t).y(t) dt  \nonumber \\
&& \quad \quad\quad \quad \quad\quad \quad \quad\quad \quad k=1,  \ldots , N-T \label{integral_inequality}  \\
&& \rho \geq 0 
\end{eqnarray}
where $j = k+T$ for some fixed $T$. The constraint of positive $\rho$ is included to make the problem infeasible if the system is not passive.

Observe that there are $2N-T+1$ constraints. The parameter $T$ is a user's choice. Smaller $T$ seems to use
more information and thus potentially result in more precise identification.
On the other hand, system noise is likely to ``distort'' the inequality \eqref{integral_inequality}
so that with noisy signals it might not be satisfied even with a passive system and a correct storage function.
Clearly, this deleterious effect of noise is reduced by increasing the window size, and eventually eliminated for sufficiently
large $T$, larger noise requiring larger $T$.
Moreover, the integrals in \eqref{integral_inequality} must be approximated numerically from the data, which is an additional
source of error that may cause the problem to be unfeasible even when the data come from a passive system,
regardless of noise.  
For these two reasons the most natural and at first glance most precise choice $T=1$ is likely to fail in most cases, except
with noise-free data, and this is what has been observed in case studies.

\section{Case study}\label{sec:case}

\subsection{The system}

Consider the following model of a pendulum
\begin{eqnarray}
\dot{x}_1 & = & x_2 \label{position} \\
\dot{x}_2 & = & - b_1 \sin(x_1) - b_2 x_2 + u \label{speed} \\
y & = & x_2 \label{saida}
\end{eqnarray}
where $x_1$ is the angular position of the pendulum with respect to the vertical and $u$ is an externally applied torque. 
This simple model is a traditional textbook example to illustrate various concepts in nonlinear systems theory - see in \cite{Khalil},
for example - but it is also a model with much practical significance in various applications, from robotics to power systems
\cite{Sauer}. In this paper, numerical results for a system with $b_1=8$, $b_2=0.5$ are presented.

Though finding a storage function from a model is not in a general an easy task, for the model \eqref{position}-\eqref{speed}-\eqref{saida} it is. 
Consider the following candidate storage function:
\begin{equation}\label{S_teorica}
S(\mathbf{x}) = \frac{1}{2}x_2^2 + b_1 (1-\cos(x_1)) .
\end{equation}
Its derivative along the trajectories of the system is given by $\dot{S}(\mathbf{x}) = - b_2 x_2^2 + x_2 .u = y.u - b_2 y^2$.
Hence the system \eqref{position}-\eqref{speed}-\eqref{saida} is $OFP(b_2)$ with storage function given by $S(\cdot )$ in \eqref{S_teorica}.
This known storage function will be used for comparison with the estimates to be obtained in the sequel.

\subsection{Identification of the storage function}

The model \eqref{position}-\eqref{speed} is used only to generate the data through its simulation,
and its knowledge is not used to identify a storage function. 
The storage function is obtained strictly from the data. The identification starts with the choice of a dictionary and
the following dictionary has been chosen:
$$
\phi (\mathbf{x})= \left[ \begin{array}{c}
x_1^2  \\
x_1x_2 \\
x_2^2 \\
(e^{x_1}-1)^2 \\
(e^{x_2}-1)^2 \\
\sin^2(x_1) \\
\sin^2(x_2) \\
(1-\cos(x_1)) \\
(1-\cos(x_2)) 
\end{array}\right].
$$

Each function in this dictionary has been included for being of a different nature - polynomial, exponential, trigonometric - and
positive definite, except the second term. Observe, however, that it composes, along
with the first and the third terms, a quadratic form, that is:
\begin{equation}\label{matriz}
\theta_1 x_1^2 +\theta_2 x_1x_2 + \theta_3 x_2^2 =  \left[\begin{array}{cc}
											x_1 & x_2
											\end{array}\right]
											P
											\left[\begin{array}{c}
											x_1 \\ x_2
											\end{array}\right] 
\end{equation}
with a rather obvious definition of the matrix $P$. Since the storage function must be positive definite, 
it is reasonable to constraint the parameters such that each piece of the dictionary is also positive, which 
is achieved including the constraints
\begin{eqnarray}
&& P\succeq 0 \label{C1} \\
&& \theta_i \geq 0 ~ ~ i= 3, \ldots, 9 . \label{C2}
\end{eqnarray}
The inclusion of these constraints to the program has also been tested.

\begin{figure}[!h]
    \centering
     \includegraphics[width=1\columnwidth]{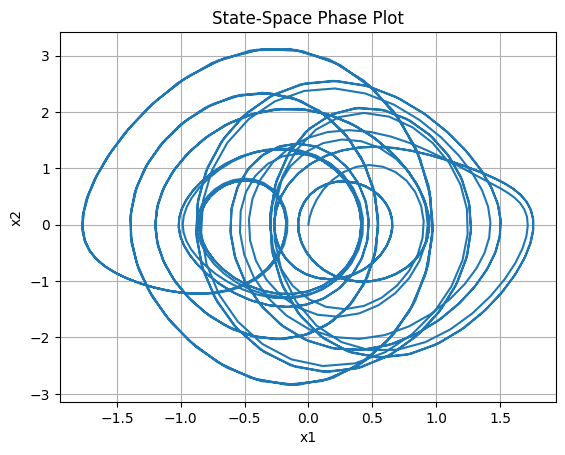}
    \caption{Phase-plot of the system that form the data set}
    \label{fig:phase_plot}
\end{figure}

Results are presented next for one particular experimental condition, defined by zero initial state and 
the following input signal:
\begin{equation}\label{entrada}
u(t) = 2(2\sin(0.2t)+\sin(t)+ \sin(2t)) .
\end{equation}
Simulations have been run for $100$ seconds and data have been collected periodically at a rate of ten samples per second; thus $N=1,000$
and the number of constraints is $2,000-T+1$. 
The noise-free system's trajectory in this experimental condition is presented in Figure \ref{fig:phase_plot}. 
Results were obtained for two cases: noise-free measurements and with the measurement of $x_1$ being perturbed
by gaussian white noise with standard deviation $\sigma$.

%

Without measurement noise and an elementary window size $T=1$ the following parameter was estimated: 
$$
\hat{\theta} = \left[\begin{array}{c}
-5.29\times 10^{-1} \\
1.73 \times 10^{-2} \\
5.11 \times 10^{-1} \\
7.83\times 10^{-4} \\
-1.68 \times 10^{-4} \\
 -2.75 \times 10^{-1} \\
8.95 \times 10^{-3} \\
9.55 \times 10^{0} \\
-3.77 \times 10^{-2} 
 \end{array}\right]
\quad \hat{\rho} = 0.471 . 
$$

The largest element in $\theta$ is $\theta_8=9.55$ and only three other elements have magnitudes larger than $|\theta_8 | / 100$.
The remaining parameters' contribution to the estimate are negligible and thus can - and should - be discarded to obtain 
a more parsimonious expression. In so doing, we end up with the estimate
\begin{equation}\label{poi}
\hat{S}(\mathbf{x}) = -0.529 x_1^2 + 0.511 x_2^2- 0.275 \sin^2(x_1) +9.55 (1-\cos(x_1)) 
\end{equation}
where only four out of the nine features present in the dictionary have been kept. 

Given that a storage function is supposed to be positive semi-definite,
it seems odd to have negative definite terms in the storage function, like the first and third terms in \eqref{poi}.
It is thus reasonable to add to the program the previously mentioned constraints \eqref{C1}\eqref{C2} 
in order to avoid that.
With these constraints in place, the optimization to be run is defined by equations
\eqref{cost}\eqref{positive_inequality}\eqref{integral_inequality}\eqref{C1}\eqref{C2} and the new parameter estimate is
$$
\hat{\theta} = \left[\begin{array}{c}
1.56\times 10^{-4} \\
1.75 \times 10^{-2} \\
4.92\times 10^{-1} \\
2.99\times 10^{-3} \\
1.27\times 10^{-4} \\
 1.54\times 10^{-2} \\
3.00\times 10^{-3} \\
7.92\times 10^{0} \\
1.12\times 10^{-2} 
 \end{array}\right]
\quad  \hat{\rho} = 0.472 .
$$
The estimate of the storage function, obtained by discarding the irrelevant features with the same $1\%$ criterion above, is
$$
\hat{S}(\mathbf{x}) = 0.492 x_2^2 +7.92 (1-\cos(x_1)) .
$$
This function is very close to the theoretical function \eqref{S_teorica}, and so is the corresponding estimate of excess of passivity.

Now let us see what happens when the measurements are noisy. In this case it is expected that 
the optimization is not feasible for small window sizes. Indeed, 
the optimization with window size $T=1$ was feasible only for $\sigma< 5\times 10^{-4}$. 
Further results will be given for  $\sigma=0.01$. 

Fixing the noise level at $\sigma=0.01$, feasibility is obtained only for $T\geq 9$. 
Even though the results are feasible and an estimate of the storage function is obtained
for window sizes $T=9$ and above, the results obtained in this range of $T$ are not satisfactory,
exhibiting high variance. For instance, here are two random
results obtained with different noise realizations for $T=10$:
$$
\hat{\theta}^{1} = \left[\begin{array}{c}
1.64\times 10^{-3} \\
5.05\times 10^{-2} \\
3.89\times 10^{-1} \\
1.01\times 10^{-4} \\
1.32\times 10^{-3} \\
 7.02\times 10^{-2} \\
1.30\times 10^{-1} \\
7.94\times 10^{0} \\
9.98\times 10^{-5} 
 \end{array}\right]
\quad  \hat{\rho}^{1} = 0.158 
$$
$$
\hat{S}_1(\mathbf{x}) = 0.389 x_2^2 + 0.130 \sin^2(x_2) + 7.94 (1-\cos(x_1)) 
$$

$$
\hat{\theta}^{2} = \left[\begin{array}{c}
1.40\times 10^{-3} \\
4.23\times 10^{-2} \\
3.23\times 10^{-1} \\
9.96\times 10^{-5} \\
4.27\times 10^{-3} \\
 1.01\times 10^{-2} \\
9.86\times 10^{-5} \\
7.60\times 10^{0} \\
3.39\times 10^{-1} 
 \end{array}\right]
\quad  \hat{\rho}^{2} = 0.223 
$$
$$
\hat{S}_2(\mathbf{x}) = 0.323 x_2^2 +  7.60 (1-\cos(x_1)) +0.339 (1-\cos(x_2)).
$$
One sees a considerable difference in the parameter values, resulting in different features being kept in the estimate of
the storage function. Moreover, the estimates of the excess of passivity in each case are very different between them and
very different from the correct value of $0.5$.

As the window size is increased, the results become more robust. 
A sample result  for $\sigma=0.01$ with positivity constraints and $T=200$ is given below:
$$
\hat{\theta} = \left[\begin{array}{c}
1.23\times 10^{-1} \\
-5.49\times 10^{-3} \\
4.94\times 10^{-1} \\
2.93\times 10^{-5} \\
3.30\times 10^{-4} \\
3.04 \times 10^{-4} \\
1.66 \times 10^{-2} \\
7.67\times 10^{0} \\
3.81\times 10^{-5} 
 \end{array}\right]
\quad  \hat{\rho} = 0.496 
$$ 
\begin{equation}\label{the_best_S}
\hat{S}(\mathbf{x}) = 0.123 x_1^2 + 0.494 x_2^2 +7.67 (1-\cos(x_1)) .
\end{equation}
Twenty Monte Carlo runs were performed for $T=200$ and $\sigma=0.01$.
In all cases the resulting estimate of passivity excess was inside the interval $0.494 \leq \hat{\rho} \leq 0.496$.
Also, in all cases only the three dictionary features present in \eqref{the_best_S} - $x_1^2$, $x_2^2$, $1-\cos(x_1)$ - were 
kept in the function's estimate (still under the $1\%$ discard  criterion),
with only the first one ($x_1^2$) being absent in some cases.
These are robust results, and are also very close to the ones obtained with the analytical  storage function \eqref{S_teorica},
despite the fact that a significant level of noise is present in the data. 

Besides the cases discussed above, simulations have also been run with various different initial conditions
and input signals, with similar results.
With the na\"ive choice of window size $T=1$ the problem is infeasible in all cases except for very low levels of noise and
in some experimental conditions the problem was infeasible even without noise, due to the errors
in the evaluation of the integrals. On the other hand, for all experimental 
conditions tested, there is always a large enough window size that provides, even with significant
levels of noise, a feasible and effective solution. 



\section{Using the storage function}\label{sec:use}

Once a storage/Lyapunov function is available, various tasks can be performed that require only its knowledge - no model needed, that is.
Among these are the determination of stability of feedback connections, estimation of a domain of attraction of the system's equilibrium 
and damping control design; these are discussed in the following Subsections.

\subsection{Feedback certification}

A classical problem in systems' theory is the study of stability of feedback connections of two or more systems. 
When the two systems being connected are a plant and its controller this is usually referred to as controller {\em certification}.


It is a known result in dissipativity  theory that if a system is $OFP(\rho )$ then its feedback connection with another system
will be stable if the other system is $IFP(\nu)$ with $\nu > -\rho$ \cite{Rodolphe}. This property allows to study the stability of
a complex system by decomposing it into a connection of several smaller systems and studying - and/or enforcing - passivity of
each component. This is by no means new, and the relevant message here is that this analysis does not require models for any of the
systems involved; only knowledge of the storage functions is required. 

%
%

\subsection{Estimating $L_fS(\mathbf{x})$}

For some purposes one also needs to estimate the derivative of the storage function along the system's trajectories.  
The estimate procedure presented above provides directly only the storage function itself and it is not obvious how to 
obtain the derivative from the data, so let me show one way to do that.

The derivative of the storage/Lyapunov function is given by
\begin{equation}\label{dotS}
\dot{S}(\mathbf{x}(t)) = L_fS(\mathbf{x}(t)) + L_gS(\mathbf{x}(t)) . u(t) .
\end{equation}

So, once $S(\mathbf{x}(t))$ has been estimated from the data, one can estimate also its derivative from  data as
$$
\dot{S}(\mathbf{x}(t)) \approx \frac{S(\mathbf{x}(t+T_s)-S(\mathbf{x}(t))}{T_s} 
$$
where $T_s$ is the sampling used in collecting the data. 
Actually for most purposes one needs $L_fS(\mathbf{x}(t))$, which is the time derivative of the storage function
along the trajectories of the autonomous system. This can be calculated directly from \eqref{dotS}
using data originated from autonomous trajectories of the system  (i.e., with $u(t) \equiv 0$)
by a procedure very similar to the one presented in Section 
\ref{sec:formulation}. But it can also be obtained from nonautonomous trajectories provided that $g(\mathbf{x})$ is known.
Indeed,
\begin{eqnarray}
L_fS(\mathbf{x}(t)) & = & \dot{S}(\mathbf{x}(t)) - L_gS(\mathbf{x}) . u(t) \nonumber \\
&  \approx & \frac{S(\mathbf{x}(t+T_s)-S(\mathbf{x}(t))}{T_s} - \nonumber \\ && \frac{L_gS(\mathbf{x}(t+T_s)) . u(t+T_s)+L_gS(\mathbf{x}(t)) . u(t)}{2} \nonumber \\
&& \label{estimate_derivative}
\end{eqnarray}
where I have taken the average value of $L_gS . u$ between two samples for a better approximation. 
Since $L_gS(\mathbf{x}(t)) = \frac{\partial S(\mathbf{x})}{\partial \mathbf{x}}. g(\mathbf{x})$, if $g(\mathbf{x})$ is known then
all the quantities in the RHS of \eqref{estimate_derivative} are known after $S(\mathbf{x})$ has been estimated.
 

\subsection{Estimate of DoA}

The domain of attraction (DoA) of an equilibrium is defined as the set of all initial conditions
which result in trajectories that converge asymptotically to it; 
in mathematical terms, if the equilibrium is the origin then the DoA is a set $\cal A$ defined as:
$$
{\cal A} = \{ \eta : \vecx(0) = \eta \rightarrow \lim_{t\rightarrow \infty} \vecx(t) = \mathbf{0} \} .
$$

As observed previously in this paper, the storage function is also a Lyapunov function for the autonomous system - that is, for $u(t)\equiv 0$.
In the possession of a Lyapunov function, estimates $\hat{\cal A}$ can be obtained for the domain of attraction of the origin. Constructing these 
estimates requires knowledge only of the Lyapunov function and its derivative, and not on the system's model, as described in the sequel.

Given a positive definite Lyapunov function $S(\vecx )$, with negative definite time derivative along the system's (autonomous) trajectories,
define the sets ${\cal D}^-$ and ${\cal L}_S^c$ as follows. The set ${\cal D}^-$ is the largest connected set containing
the origin such that $L_fS(\vecx ) < 0 \forall \vecx\in D^- / \{\mathbf{0} \}$. The sets ${\cal L}_S^c$ are the level sets of the Lyapunov function, whose boundaries
are defined  by
$$
\partial {\cal L}_S^c = \{ \vecx : S(\vecx ) = c \}
$$
for some $c>0$. The core of Lyapunov theory is that all level sets contained in ${\cal D}^-$ are positively invariant.
The most immediate and most important corollary of this property is that
any bounded level set inside ${\cal D}^-$ is contained in the DoA of the origin. 
The best possible conservative estimate (that is, such that ${\hat{\cal A}} \subseteq {\cal A}$) of the DoA is thus given
by ${\hat{\cal A}} = {\cal L}_S^{\bar{c}}$ where $\bar{c}$ is the largest value of $c$ for which ${\cal L}_S^c$ is bounded and ${\cal L}_S^c\subseteq {\cal D}^-$.

\begin{figure}[!h]
    \centering
 \includegraphics[width=1\columnwidth]{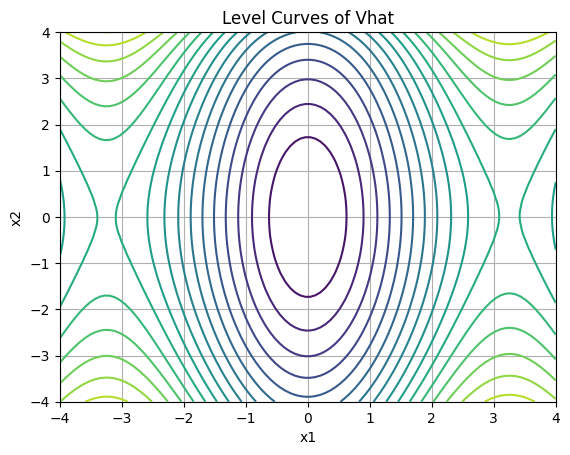}
    \caption{Levels curves of the estimated Storage function $\hat{S}(\vecx )$}
    \label{fig:levels}
\end{figure}

The level sets ${\cal L}_S^c$ of the estimated Lyapunov function \eqref{the_best_S} are plotted in Figure \ref{fig:levels}.
As for the negative set ${\cal D}^-$, by using  \eqref{estimate_derivative} with the previous data one finds that $L_fS (\mathbf{x})<0 $
for all points in that trajectory. Hence ${\cal D}^-$ contains the whole region inside the system's trajectory shown in
Figure \ref{fig:phase_plot}. With the available data, this is the largest estimate that can be made for ${\cal D}^-$, so I will consider
that is the set inside the trajectories in Figure \ref{fig:phase_plot}.
The third level curve in Figure \ref{fig:levels} is the largest one that fits within the set thus defined. 
So, this is the best estimate of the DoA that can be obtained with this Lyapunov function and with these data. One can write
the estimate analytically as (from (22))
$$
\hat{\cal A} = \{ \vecx :    0.123 x_1^2 + 0.494 x_2^2 +7.67 (1-\cos(x_1)) < 4.45 \}
$$


This is actually not a tight estimate of the real DoA, but the data are to ``blame'' for it.
The available data only cover the region shown in Figure \ref{fig:phase_plot}, so they do not
tell anything about the behavior of the system outside this region. 
Actually this is a very important point to be made about data-driven analysis and design:
using only data one can never safely extrapolate outside the domain of the state space that is visited by the data. 
Meaningful extrapolation requires some prior. For instance, should we somehow know as a prior that the real
set ${\cal D}^-$ is the whole state space (which is the case here) then the estimate would be given by the largest
bounded level curve in Figure \ref{fig:levels}, which is quite a tight estimate.


\subsection{Damping control}

Given a storage function $S(\mathbf{x})$, its derivative along the trajectories of the system is given by
$$
\dot{S}(\mathbf{x}) =  L_fS(\mathbf{x}) + L_gS(\mathbf{x}) . u .
$$
So, if a controller is designed as $u=-k. L_gS(\mathbf{x}) $, then for any $k>0$ the derivative is more negative in closed-loop
than it was in open-loop and the same function is still a Lyapunov function for the closed-loop system. Such a control law
is usually called $L_gV$ control, or damping control.

Now, to design such a controller, again one does not need to know the model of the system, as the 
control law is given by 
\begin{equation}\label{LgVcontrol}
u = -k \frac{\partial S(\mathbf{x})}{\partial \mathbf{x}} g(\mathbf{x}),
\end{equation}
which does not depend on the vector field $f(\cdot )$, only on $g(\mathbf{x})$.
But knowledge of the function $g(\mathbf{x} )$ is not necessary either. Instead of \eqref{LgVcontrol}, one can
get the same effect with the control law
\begin{equation}\label{LgVcontrol_sorta}
u = -k \frac{\partial S(\mathbf{x})}{\partial \mathbf{x}} b
\end{equation}
where $b\in\Re^n$ is a constant vector where each nonzero element of $g(\mathbf{x})$ is replaced by a constant value
with the same signal as the replaced function. 
In other words, to design the damping controller one needs to know only at which state variables the input enters and 
in  what sense - with positive or negative gain. In the example, $g(\mathbf{x})= [0 ~1]^T$, so an $L_gV$ controllers is 
$u = -k.x_2= - k.y$. 

\section{Conclusion}\label{sec:conclusion}

In this paper I presented a purely data-driven method to analyze the passivity of a system. 
The results obtained for a case study (the pendulum benchmark)  are promising.
This method is useful even when a precise model for the system is available, since obtaining a
good storage function from a model is not always possible. On the other hand, when a model (or even
the sketch of one) is available, the proposed procedure
is likely to be more effective, because then it will 
be easier to come up with a good dictionary.

I have also stressed that knowledge of the storage function is enough to perform various tasks.
Combined with some scarce knowledge about the system's model, like knowing the input vector
$g(\cdot )$, or only its $0/1$ structure, additional tasks can be performed.     
Future (and present) work involves a theoretical analysis of the effects of noise, the development
of other control methods that will require knowledge only of the storage function, and also testing the procedure in more complex systems.  

\bibliographystyle{plain}        
\bibliography{passivity}

\end{document}